\documentclass[prl,twocolumn,aps,showpacs,superscriptaddress]{revtex4-1}
\usepackage{graphics}
\usepackage{epsfig}
\usepackage{amsmath, bm}
\usepackage{amssymb}
\usepackage{color}
\usepackage[utf8x]{inputenc}

\graphicspath{{Figures/}}
\newcommand{\rom}[1]{\uppercase\expandafter{\romannumeral #1\relax}}

\begin{document}
\newcommand{\beq}{\begin{equation}}
\newcommand{\eeq}{\end{equation}}

\title{Design of pseudo-mechanisms and multistable units for mechanical metamaterials}

\author{Nitin Singh}
\affiliation{AMOLF, Science Park 104, 1098 XG Amsterdam, The Netherlands}
\author{Martin van Hecke}
\affiliation{AMOLF, Science Park 104, 1098 XG Amsterdam, The Netherlands}\affiliation{Huygens-Kamerlingh Onnes Lab, Leiden University, PObox 9504, 2300 RA Leiden, The Netherlands}

\date{\today}

\begin{abstract}
Mechanism---collections of rigid elements coupled by perfect hinges which exhibit a zero-energy motion---motivate the design of a variety of mechanical metamaterials.
We enlarge this design space
by considering {\em pseudo-mechanisms}, collections of elastically coupled elements that exhibit motions with very low energy costs. We
show that their geometric design generally is distinct from those of true mechanisms, thus opening up a large and virtually unexplored design space.
We further extend this space by designing building blocks with bistable and tristable energy landscapes, realize these by 3D printing, and show how these form unit cells for multistable metamaterials.
\end{abstract}

\pacs{81.05.Xj, 81.05.Zx, 45.80.+r, 46.70.-p}

\maketitle

A mechanism is a collection of flexibly linked, rigid elements which exhibits a zero-energy motion. Mechanisms play a foundational role in the physics of jammed media and spring networks \cite{alexander,network1,network2,network3,network4,jam1,jam2}, and are central in
mechanical engineering, where they underlie the design of robotic devices such as grippers \cite{robot,Detweiler}. Imperfect mechanisms, based on distorted geometries \cite{Dudte:2016db,Jesse,Stern:2017bya,Meeussen}, extra bonds/connections \cite{alexander,network1,network2,network3,network4,jam1,jam2}, or non-ideal hinges \cite{natphyschain} frequently occur, and these exhibit soft modes similar to the zero energy motions of the underlying mechanism.
In particular, mechanism-based metamaterials borrow the geometric design of mechanisms but replace the hinges by slender, flexible parts which connect stiffer elements, such that the soft modes of the metamaterial are similar to the free motion of the underlying mechanism.
External forces easily excite these soft modes, and as the mechanism-derived soft modes can be very different from those of ordinary elastic modes,
exotic properties may emerge \cite{review}, including negative response parameters \cite{auxetic,katia,Reid}, shape-morphing \cite{cube,Overvelde:2017,Chiara,kimnatphys}, topological polarization \cite{Jayson,Chen:2016bk,Marc,zeb}, programmability and multistability \cite{Jesse, bastiaan,
Waitukaitis:2015dk,Dudte:2016db,rafsanjani} and (self-)folding \cite{Overvelde:2017,Chen:2016bk,Jesse,Waitukaitis:2015dk,Dudte:2016db,Filipov:2015,Stern:2017bya,Filipov:2017,sequential,Stern:2018,Chen:2019,natphysori}.



However, as mechanism-based metamaterials do not have true zero modes \cite{review,natphyschain},
the design of a flexible metamaterial does not require an underlying true zero-energy mechanism. This suggests to consider
{\em pseudo-mechanisms} (PMs), which we define as collections of flexibly coupled rigid elements that exhibit motions with (very) low energy costs.





Here we show that PMs are widespread, by constructing a  quadrilateral based PMs by use of particle swarm optimization.
Our central finding is that most PMs are geometrically very distinct from true mechanisms; most PMs are not simply perturbed mechanisms, but PMs permeate the design space very far
away from the true mechanism subspace.
We extend our search techniques to obtain multistable units \cite{rafsanjani}, and bring these to life using 3D printing. Finally, we show how to tile our unit cells to obtain complex periodic metamaterials. Together, our approach, which is computationally effective, suggests new avenues for the design of shapemorphing and multistable metamaterials \cite{review} as well as devices for robotics or deployable structures such as bellows \cite{robot,deploy}.

\begin{figure}[t]
\begin{center}
	\includegraphics[width=\columnwidth,bb=23 0 325 120,clip]{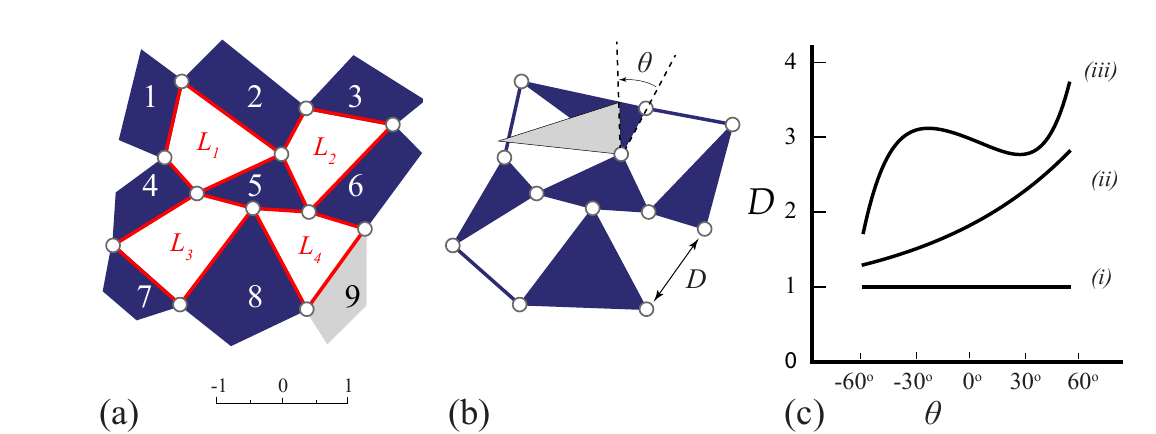}
\end{center}
	\caption{(color online) (a) Full unit consisting of nine rigid quadrilaterals (1-9), connected by twelve flexible hinges $\{x_{12},x_{23},\dots \}$ (circles). This unit can also be seen as four connected four-bar linkages $L_i$.
(b) Diluted unit.
(c) Depending on the design of quadrilaterals (1-8), the characteristic motion of the diluted unit, $D(\theta)$, can be (nearly) constant \textit{(i)}, a monotonic function \textit{(ii)} or a non-monotonic function \textit{(iii)}. }
	\label{fig:1}
\end{figure}

{\em System.---}
We consider collections of quadrilaterals connected in a square topology by hinges with zero torsional stiffness and finite stretchability (Fig.~1). For equally sized squares, such system has a zero-mode and
is known as the rotating square mechanism \cite{grima}; this geometry
underlies a large number of mechanical metamaterials \cite{review,katia,cube,Chiara,bastiaan,sequential,natphyschain,luuk}. Generalizations, including to regular tilings of alternatingly sized squares, rectangles or 3D, are well known \cite{cube,bastiaan,Finish_dude,kirisq}.
The condition for such collections of quadrilaterals to form a mechanism are simple.
For definiteness, we focus on $3 \times 3$ tilings of quadrilaterals (Fig.~1a).
We can consider such tilings as collections of connected four-bar linkages $L_i$, and then express the relations between their (inner) angles by mappings $M_i$.
It can be shown that quadrilateral tilings can only form a mechanism if all four-bar linkages (voids) form parallelograms, as these are associated with linear mappings $M_i$ \cite{geom}.
In contrast, for generic quadrilaterals the mappings $M_i$ are nonlinear, and
tilings of $3\times 3$ (or larger) generic quadrilaterals do not poses a zero energy motion \cite{geom,luuk,MC} (Fig.~1a).

To make progress, we focus on a diluted unit, obtained
by removing quadrilateral 9, which yields
a {\em mechanism} with a freely hinging, zero energy, finite amplitude mode \cite{luuk,MC} (Fig.~1b).
To characterize this mechanism, we remove all extraneous information, and replace the corner quadrilaterials with rigid bars (Fig.~1b).
The geometry of this mechanism is specified by the coordinates of the 12 links
$\{x_{12},x_{23},\dots\}=:X$, which span a 24 dimensional design space.
We control the
free motion of this mechanism by $\theta$, the deviation of $\angle $ from its initial value, and characterize the diluted unit by $D:=|x_{89}-x_{69}|$ as function of $\theta$ (Fig.~1b).
Experimentally, stretching or compressing two points on the systems, or compressing it between parallel plates actuates the soft mode of the system that we describe here.
The function $D(\theta)$ acts as a proxy for the mechanics of a full $3 \times 3$ unit consisting of flexible elements: if $D(\theta)$ is a constant, reinserting a ninth quad of appropriate dimensions yield a full $3 \times 3$ unit with a zero mode \cite{MC,note1}
Fig.~1c(\textit{i}). For nearly constant $D(\theta)$, reinserting the ninth quadrilateral
would lead to a system with a large amplitude motion with a very low energy:  a {\em pseudo-mechanism}. For generic quadrilaterals  $D(\theta)$ is a nonlinear function (Fig.~1c(\textit{ii}-\textit{iii})), and inserting the ninth quadrilateral yields a more complex energy landscape.
The design problem is thus to obtain
coordinates $\{x_{12},x_{23},\dots\}$ so that $D(\theta)$
closely matches a target function $D_t(\theta)$.

\begin{figure}[t]
	\includegraphics[scale=0.70]{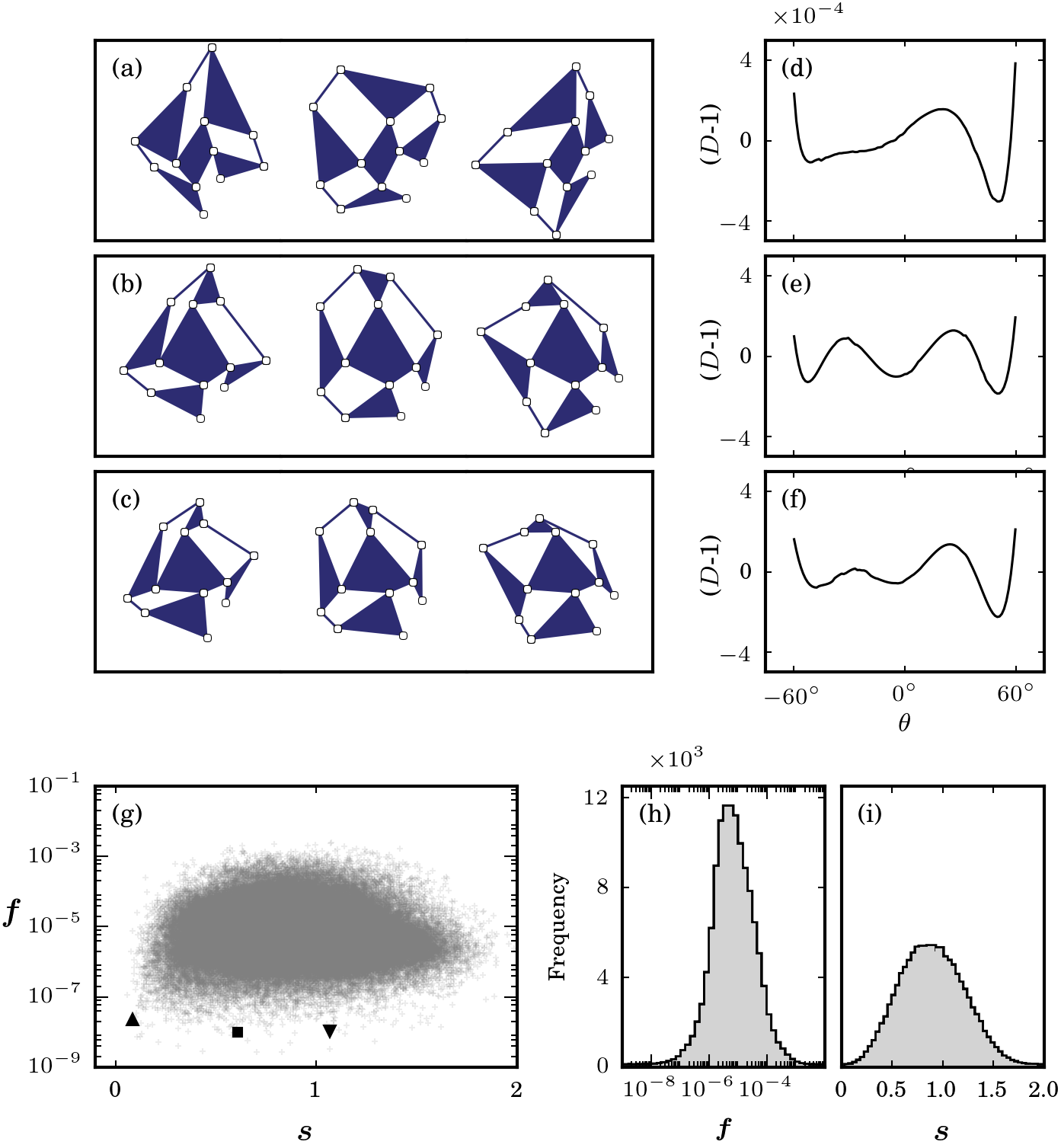}
	\caption{(a-c) Three examples of diluted units for which $D(\theta)$
 is nearly constant and equal to 1, for $\theta\! \in\! \left[-60^{\circ},60^{\circ}\right]$, for $\left[f,s\right] \!=\!\left[2.26\times10^{-8},0.084\right]$ (a); $\left[9.88\times 10^{-9},0.611\right]$ (b); and $\left[1.04\times10^{-8},1.06\right]$ (c).
The three snapshots in each panel correspond to  $\theta\!=\! -60^{\circ}$, $0^{\circ}$ and $60^{\circ}$.
(d-f) Corresponding plots of $D-1$ vs $\theta$; notice the scale. (g)
Scatter-plot of $f$ vs $s$, where uptriangle, square and downtriangle symbols indicate the parameter values shown in panels (a-c) respectively. (h,i) Distributions of $f$ and $s$.}
	\label{fig:2}
\end{figure}

{\em Design by particle swarm optimization.--- }
We define a cost function $f$
based on the normalized Euclidean distance between $D(\theta)$ and
$D_t(\theta)$, combined with discrete constraints to avoid
non-fitting quadrilaterals, overlapping quadrilaterals, and designs where the quadrilateral sizes differ too much (see S.I.).
Exploring this design space requires an algorithm that does not easily get stuck in shallow minima, as purely gradient based methods would. Evolutionary algorithms are eminently suited for this, and we choose here to use particle swarm optimization (PSO) due to its simplicity and ease of tuning. This method employs an ensemble (swarm) of particles - each representing a particular design - and is known to allow to identify deep minima in a rugged landscape \cite{pso1,pso2,pso3,pso4,pso5,pso6,pso7}.
While we note that our approach remains valid for larger structures, the computational time grows exponentially in the size of the structure and we focus here on $3\times3$ structures.
The PSO algorithm keeps track of the best position discovered by each particle up to generation (iteration) $k$, ${\bf xb}_i^k$, and  by the best position discovered by all the particles --- the swarm --- ${\bf xs}^k$. We seed
an initial population of randomly distributed particles with random velocities.
During the search, each particle is attracted towards a stochastic mix of ${\bf xb}_i^k$ and ${\bf xs}^k$:
\begin{eqnarray}
{\bf v}_{i}^{k+1} &\!=\!& w {\bf v}_{i}^{k} \!+\! c_1  {\bf r}_i^1 \cdot ({\bf xb}_{i}^k \!-\! {\bf x}_{i}^k) \!+\! c_2 {\bf r}_i^2\cdot ({\bf xs}^k \!-\! {\bf x}_{i}^k~), \\
{\bf x}_{i}^{k+1} &\!=\!& {\bf x}_{i}^{k} \!+\! {\bf v}_{i}^{k+1}~.
\end{eqnarray}
where ${\bf r}_i^1$ and ${\bf r}_i^2$ are random vectors whose elements are uniformly distributed between 0 and 1, and the so-called inertia
$(w)$, cognition $(c_1)$ and social $(c_2)$ hyper-parameters must be chosen to optimize convergence. For our specific design problem,
the position ${\bf x}_i$ and velocity ${\bf v}_i$ of particle \textit{i} are both 24-dimensional vectors, and we have verified by hyper-parameter optimization that the algorithm yields good results for $w=0.25$, $c_1 \geq 0.0$, $c_2 \geq 1.75$ and $c_1+c_2 \leq 3.50$ (see SI).
For each target function, we
run 3000 runs for each of the 36 pairs of parameter values that satisfy
$c_1=0,0.25,0.5,\dots,1.75$, $c_2=1.75,2,2.25,\dots,3.5$ and $c_1+c_2 \leq 3.5$, leading to a total of $1.18\times10^5$ runs. For details, see the Supplemental Information.

\begin{figure*}[t]
\includegraphics[width=17.7cm]{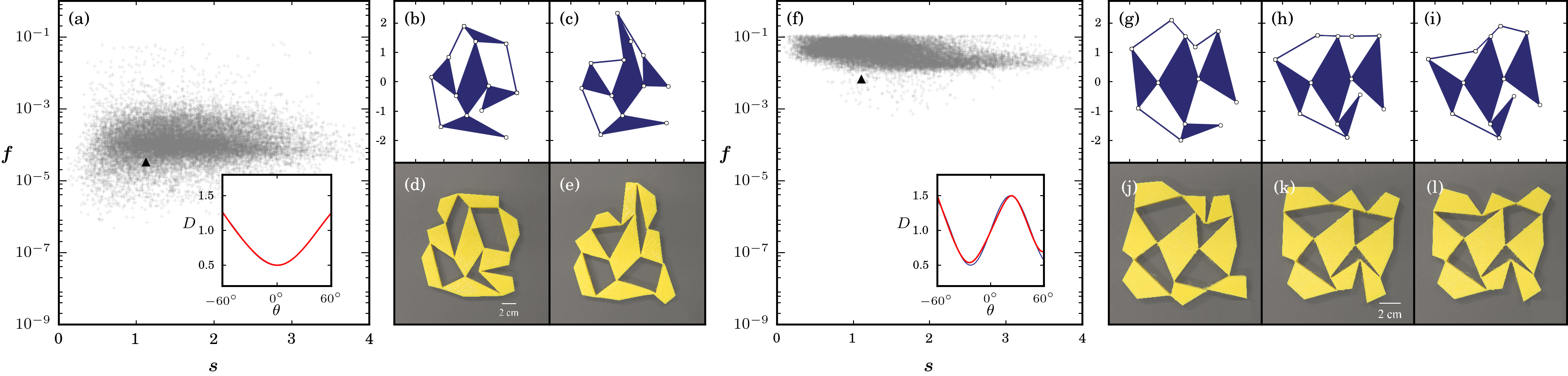}
\caption{(a) Scatter plot [18000 points] of $f$ and $s$ for the target function  $D_t(\theta)=1+0.5\sin\left(2\theta+\pi/2\right)$.
Inset: The target and {realized} $D(\theta)$
for {$f=3.2\times 10^{-5},s=1.12$ (triangle in panel (a)) are virtually indistinguishable}.
(b,c) Corresponding pruned unit for  $\theta=\pm 60^{\circ}$, where
$D\approx 1$ (d,e) Corresponding 3D printed flexible bistable unit in both its stable states.
(f) Scatter plot [18000 points] of $f$ and $s$ for a target function $D_t(\theta)=1+0.5\sin\left(4\theta\right)$.
Inset: The target and {realized} $D(\theta)$ for $f=6.7\times 10^{-3},s=1.10$ (triangle in panel (g)) are virtually indistinguishable.
(g-i) Corresponding pruned unit for  $\theta=-45^{\circ},0^{\circ},45^{\circ}$, where
$D\approx 1$.
(j-l) 3D printed tristable unit in all three stable states.}
	\label{fig:3}
\end{figure*}

{\em Generic flexible unit cells.--- }  We first focus on designing diluted units
for which $D(\theta)$ is close to a constant. We set the target curve $D_{t\mathit{1}}(\theta) = 1$ and deploy PSO to obtain designs with low values of $f$.
We find a large number of designs for which $f$ is very small, so that
$D(\theta)$ is close to a constant (Fig.~2).
We quantify the geometry of these solutions through an order-parameter, $s$, which measures the proximity of the four internal four-bar linkages to parallelogram linkages. We  define $\bm{s_i}$ for the $i^{th}$ linkage as
\begin{equation}
{s_i^2} = \dfrac{(a_{i}^1 - a_{i}^3)^2 + (a_{i}^2 - a_{i}^4)^2}{\textstyle\sqrt{(a_i^1)^2 + (a_i^2)^2 + (a_i^3)^2 + (a_i^4)^2}}~,
\end{equation}
where $a_i^j$ are the bar lengths $j=1,\dots,4$ of linkage $i$, and define $s$ as
\begin{equation}
s = \sqrt{\sum\limits_{i=1}^4 s_i^2} ~.
\end{equation}
While our algorithm finds some solutions with small $s$, i.e., close to the true mechanism limit where all linkages are parallelograms (Fig.~2a), the vast majority of solutions with low $f$ have significantly larger values of $s$ (Fig.~2b-c). Notwithstanding this strong deviation from true mechanisms, the peak deviation between $D(\theta)$ and $D_t(\theta)$ can be as small as $4\times 10^{-4}$ (Fig.~2d-f).

We show a scatter plot of
$f$ versus $s$, and the individual distributions of $f$ and $s$ --- which are only weakly correlated --- in Fig.~2g-i. These plots reveal that the distribution of $f$ is log-normal, with $s$ normally distributed with the center at $s\approx 1 $, corresponding to designs that are very far away from strict mechanisms ($s=0$). Hence, pseudo-mechanisms with anomalously low functional deviations from true mechanisms are widespread, and occur in regions of design space that are far away from true mechanisms.

Our findings suggest a complex organization of the design space.
To gain insight into this structure, we have explored whether
the value of $f$ increases if a certain solution $x_0$ is randomly perturbed. Specifically, we generate 1000 random 24 dimensional vectors $dx$ with each entry uniformly distributed between -1 and 1, and then calculate $f(x_0+\varepsilon dx)$ for a range of $\varepsilon$. For the deep {solution, where $\left[f_0,s\right]=\left[9.88\times 10^{-9},0.61\right]$}, we
find that all $f>f_0$, consistent with the idea that these solutions are local minima (see S.I.). In contrast, for solutions with much larger values of $f$ we find a small but finite probability that $f<f_0$ for small perturbations ($\varepsilon = 10^{-3}$) but not for larger perturbations of order $\varepsilon = 10^{-2}$. We suggest that these solutions perhaps are close to a shallow local minimum, and note that PSO is not guaranteed to find local minima with high accuracy  (For details, see S.I.).

{\em Multistable unit cells.--- }
The ease with which we can find pseudo mechanisms prompts the question if it is similarly easy to generate designs for other target functions.
For systems with flexible hinges, inserting a ninth quad with dimension $D'$ provides the blueprint for a unit with low energy states for $D(\theta)=D'$, so that nonmonotonic
$D(\theta)$ lead to multi-stable structures.
We have investigated four families of target functions, $D_{1,t}=1+\alpha\theta$; $D_{2,t}=1+\alpha\sin(2\theta+\pi/2)$; $D_{3,t}=1+\alpha\sin(3\theta+\pi/2)$ and $D_{4,t}=1+\alpha\sin(4\theta)$, for a range of values of $\alpha$ between $-1/2$ and $1/2$
\cite{thesisnitin}. Here we focus on the designs for $D_{2,t}$ and $D_{4,t}$, as these form the basis for bistable and tristable units.

We show scatter plots of $f$ vs $s$ for $D_{2,t}$ and $D_{4,t}$ in Fig.~3, for
$\alpha=0.5$. We observe a large cloud of solutions, and note that the typical values of $f$ for curves with more extrema are somewhat larger than those for $D_t=1$.
Examples of designs of diluted units that closely satisfy the target curves are shown in Fig.~3.

We experimentally realized full units based on the designs shown in Fig.~(3b,g),
by adding a ninth quad of appropriate length, and then 3D printing these units with flexible material (filaflex). The out-of-plane thickness of these sample is 10 mm and the connecting hinges have a minimum thickness of $\approx$ 0.5 mm. Despite the finite flexibility of all quads, and the finite but small bending stiffness of their hinges,
we observe that these samples are indeed bistable and tristable respectively, with their stable configurations close to the expected configurations (Fig.~3d,e,j,k,l).

\begin{figure}[t]
	\includegraphics[width=1\linewidth]{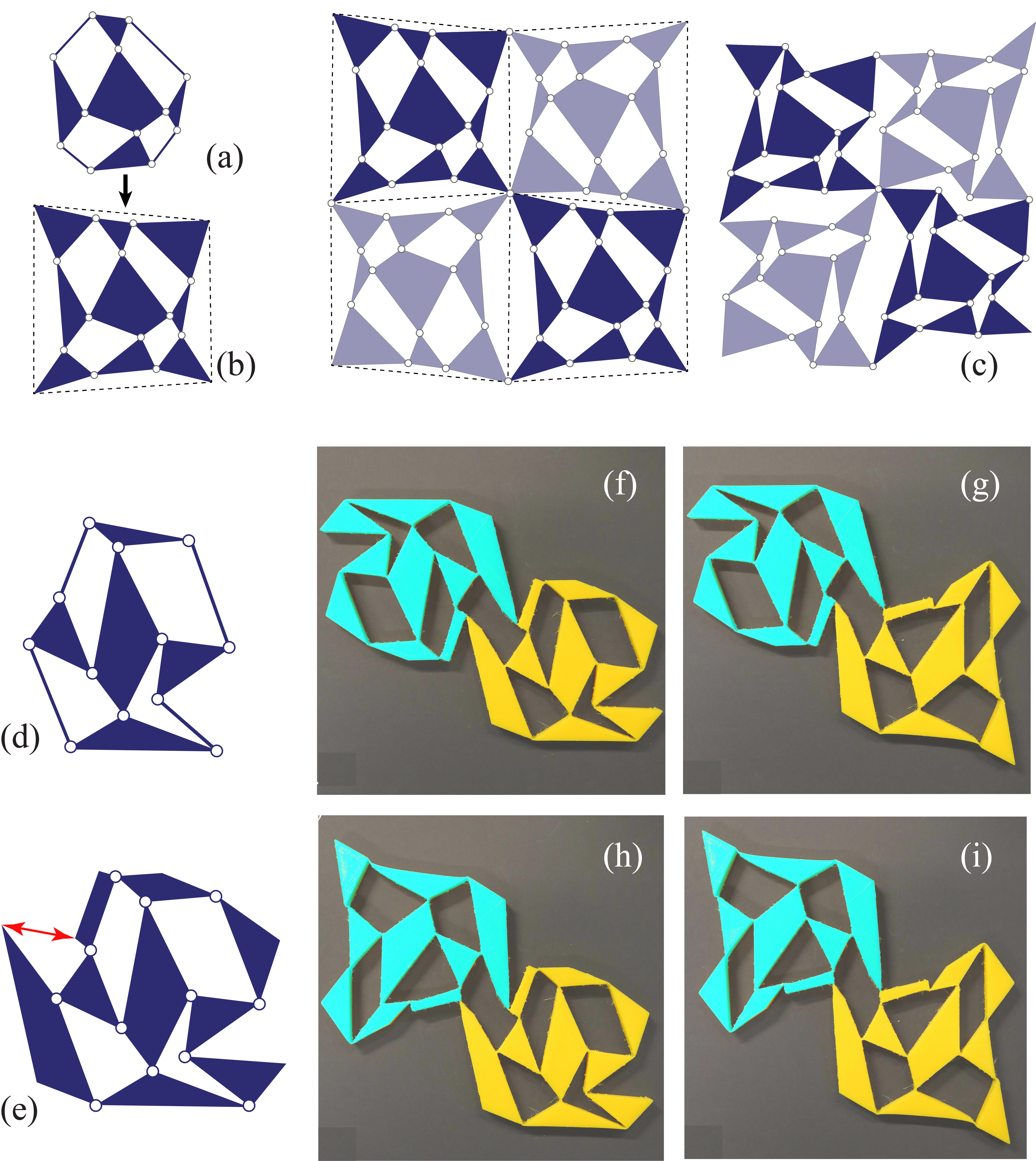}
	\caption{(Color online) (a-b) Design of a PM augmented by replacing the outer bars by triangles. (c) A staggered tiling of such PMs has a soft hinging mode.
(d) Bistable design. (e) Augmented design where gap distance (red arrow) has the same length in both stable states. (f-i) Two connected bistable 3D printed unit cells (green, yellow) can be snapped between four different stable states (false green color added for visibility).
 }
\end{figure}

{\em Complex Tilings.--- }
Finally, we briefly outline how we can connect complex $3\times 3$ units into larger systems. Each PM can be augmented by replacing the outer bars by
triangles (Fig.~4a-b). The outer tips of this unit form a quadrilateral, and
as any quadrilateral can be tiled in a pattern where adjacent quadrilaterals are rotated by $180^{\circ}$, larger PMs can readily be designed by connecting these units (Fig.~4c). One can similarly augment multistable unit cells, and for a tiling of general augmenting triangles one expects stable collective states only when all units are in the same configuration, as the `gap' distance between tips of triangles generally differs in different minima. However, the augmenting triangles can be chosen such that this gap has the same length in each stable state (Fig.~4d-e). Connecting such augmented units in a tiling yields a design with energy minima when each individual unit is in its stable state, leading to a number of stable states which grows exponentially with system size (Fig. 4f-i).

{\em Summary and Outlook.--- }
We have presented a novel strategy for the design of metamaterial architectures, based on pseudo-mechanisms which can have a geometric structure which is surprisingly far removed from that of strict mechanisms.
As similar pseudo-mechanisms can be observed in 2D origami,
where PMs allow to circumvent the difficult design of rigidly folding mechanisms \cite{Dudte:2016db,Stern:2017bya,Stern:2018,natphysori},
we speculate that pseudo-mechanisms are generic and relevant for a wide classes of structures, including networks of hinged bars \cite{kimnatphys} and (3D) origami \cite{Overvelde:2017}. Moreover, the ease of designing  multistable structures in a hierarchical fashion---coupling complex units in tilings---suggest to generalize this approach to other classes also.

Extensions of our work include the design of larger non-periodic collections of quadrilaterals that form pseudo-mechanisms. Conceptually, the step from a $2\times3$ mechanism to a $3\times 3$ pseudo mechanism might be similar to that from a $3\times 3$ to a $3\times 4$ pseudo mechanism, but it is an open question how the design space evolves for increasingly large systems. A further intriguing possibility arises for, e.g., bellows: while the volume of a polyhedron cannot change as it flexes,
pseudo-mechanisms may in practice work equally well \cite{deploy}.
Moreover, we wonder whether pseudomechanisms can mimic an equivalent of the topological polarization, edge-modes and corner-modes observed in topologically non-trivial mechanical metamaterials that are based on true mechanisms \cite{Jayson, Chen:2016bk,Marc,zeb}.
Finally, our designs space is only of moderate dimensions, and obtaining nontrivial designs  is computationally relatively cheap. This makes our designs eminently suited to test whether machine learning techniques would be suitable to, first, be trained to distinguish ``good'' from ``bad'' pseudo mechanisms, second, to detect and classify multistable designs, and third, can be used to speed up the design of such structures \cite{MLori,bessa}.

{\em Acknowledgements.---}  We thank M. Bessa, M. Dijkstra, S. Guest, A. Murugan,  S. Pellegrino and T. Tachi for  productive discussions. This work is part of an Industrial Partnership Programme of the Netherlands Organization for Scientific Research (NWO) under grant nr 12CSER036.

\bibliographystyle{apsrev}

\newpage
\clearpage

\section{Supplemental Information}


\subsection{Objective function}

We couple the rigid quads with springs of zero restlength and unit stiffness, and for given $\theta$ minimize the elastic energy with standard conjugate gradient techniques (to essentially zero) to obtain $D(\theta)$ --- this method makes it easy to deal with problems that may occur when some quadrilaterals grow too large or too small and are no longer able to connect to their neighbors. We define the objective function $f$ as the sum of the normalized Euclidean distance between $D(\theta)$ and $D_t(\theta)$ ($f$), and three constraints ($p,q,r,$): ${f}  = {g} + {p} + {q} + {r}.$
Here $g$ is defined as $1/N \Sigma_{i=1}^N (D(\theta_i)-D_t(\theta_i))^2 $, where $\theta_i$=$
-60^{\circ},-54^{\circ},-48^{\circ},\dots
60^{\circ}$.

{\em Disconnect constraint $p$.--- } When some quadrilaterals grow too large or too small and are no longer able to connect to their neighbors, the energy cannot equilibrate to zero. We have found that for our numerical precision, $E<10^{-10}$ for proper systems. We define
for each $\theta_i$ a penalty $p_i=0$ when $E<10^{-10}$, $p_i=(\log_{10}\bm{E}_{i}) / 10 + 1$ otherwise, and define $p:= \Sigma_i p_i$.

{\em Overlap constraint $q$.--- } During optimization, the evolving design variables  may result in systems where some quadrilaterals overlap during their hinging motion. We
identify such self-intersecting systems by first defining:
\textit{(i)} the outermost polygon $P_o$, defined by its corners $\{ x_{14},x_{12},x_{23},x_{36},x_{68},x_{78},x_{47}\}$,
\textit{(ii)} the `windmill-shaped' polygon $P_w$, which encloses the four linkages $L_i$ and  quadrilateral 5, and is defined by its corners $\{x_{12},x_{25},x_{23},x_{36},x_{56},x_{69},x_{89},x_{58},x_{78},x_{47},x_{45},x_{14}\}$
and
\textit{(iii)} the inner polygon $P_i$ defined by $\{x_{25},x_{56},x_{58},x_{45}\}$, i.e., quadrilateral 5.
The necessary and sufficient conditions to guarantee a non self-intersecting system are:
\textit{(i)} $P_i$ is `contained within' $P_o$, and \textit{(ii)} all three $P_o$, $P_w$ and $P_i$ are simple and do not self-intersect.

Similar as above, we need to check for the violation of the present constraint for every step of $\theta_i$. Its value at the $i^{th}$ $\theta$ step is denoted by ${q_i}$. A simple binary quantification for ${q_i}$ is implemented, where it is assigned a value 1 if self-intersection occurs and 0 otherwise.
The total violation for the complete range of $\theta$ is given by ${q}$:
\begin{equation}
{q} = \sum\limits_{i=1}^{20} q_i
\end{equation}
Finally we note that $q_{\mathit{i}}$ can only be calculated if $p_{\mathit{i}}$ = 0. For $p_{\mathit{i}} \neq 0$, $q_{\mathit{i}}$ is simply assumed to be zero, and the $p$ constraint is sufficient to suppress such solutions

{\em Size constraint $r$.--- } We occasionally observe that, driven by penalties $p$ and $q$, systems with disproportionate sizes of their quadrilaterals arise. In order to avoid such systems, we impose a third constraint whose aim is to keep every edge length of every polygon within a desired range between $0.5$ and $2.5$. For each edge $j$ with length $l_j$ of each polygon
%
we specify a minimum  length  $l_{min}=0.5$ and a maximum edge length $l_{max}=2.5$, and assign a penalty $r_j$ by a piecewise linear function:
$r_j = (1 - 0.5l_j)$ for $0 \leq l_j < 0.5$;
$r_j = 0$ for $0.5 \leq l < 2.5$;
$r_j = 0.5(1 - 0.5l_j)$ for  $2.5 \leq l_j < 3.0$;
$r_j=1 $ for  $l_j \geq 3.0$.
The total penalty $r$ is the sum
\begin{equation}
{r} = \sum\limits_{j=1} r_i.
\end{equation}

The typical magnitude of these three constraints during violations is significantly larger than the Euclidean distance between $D(\theta)$ and $D(\theta_i)$, and as a result
the collections of quadrilaterials obtained by our PSO algorithm satisfy these constraints and are bound in size, remain connected during hinging, and do not overlap.

\subsection{Particle Swarm Optimization}

Here we briefly summarize some of the technical details of our implementation of PSO.

\paragraph{Position initialization --} The search procedure begins by spreading
the particles throughout search space [48, 49].
For each PSO particle, we first place the coordinates $X$ corresponding to a rotating square
mechanism where each quadrilateral has a diagonal of length 1.5, and then perturb each of the 24 coordinates of $X$ with random numbers uniformly distributed between $-1/2$ and $1/2$. We then check whether any constraint is violated, and if so, generate a new particle, until
all particles satisfy all constraints.

{\em Velocity initialization.--- } We initialize the velocity of each particle in each dimension by a random number uniformly distributed between 0 and 1.

{\em Swarm size.--- } The number of particles in the swarm aims to strike a balance between good coverage of the search space and computational efficiency. We found that for our problem a swarm size of 50 is adequate.

{\em Termination criteria.--- } We terminate the PSO search when the number of iterations reaches 100.

\subsection{Hyper-parameter Optimization}

\begin{figure}[t]
	\includegraphics[width=1\linewidth]{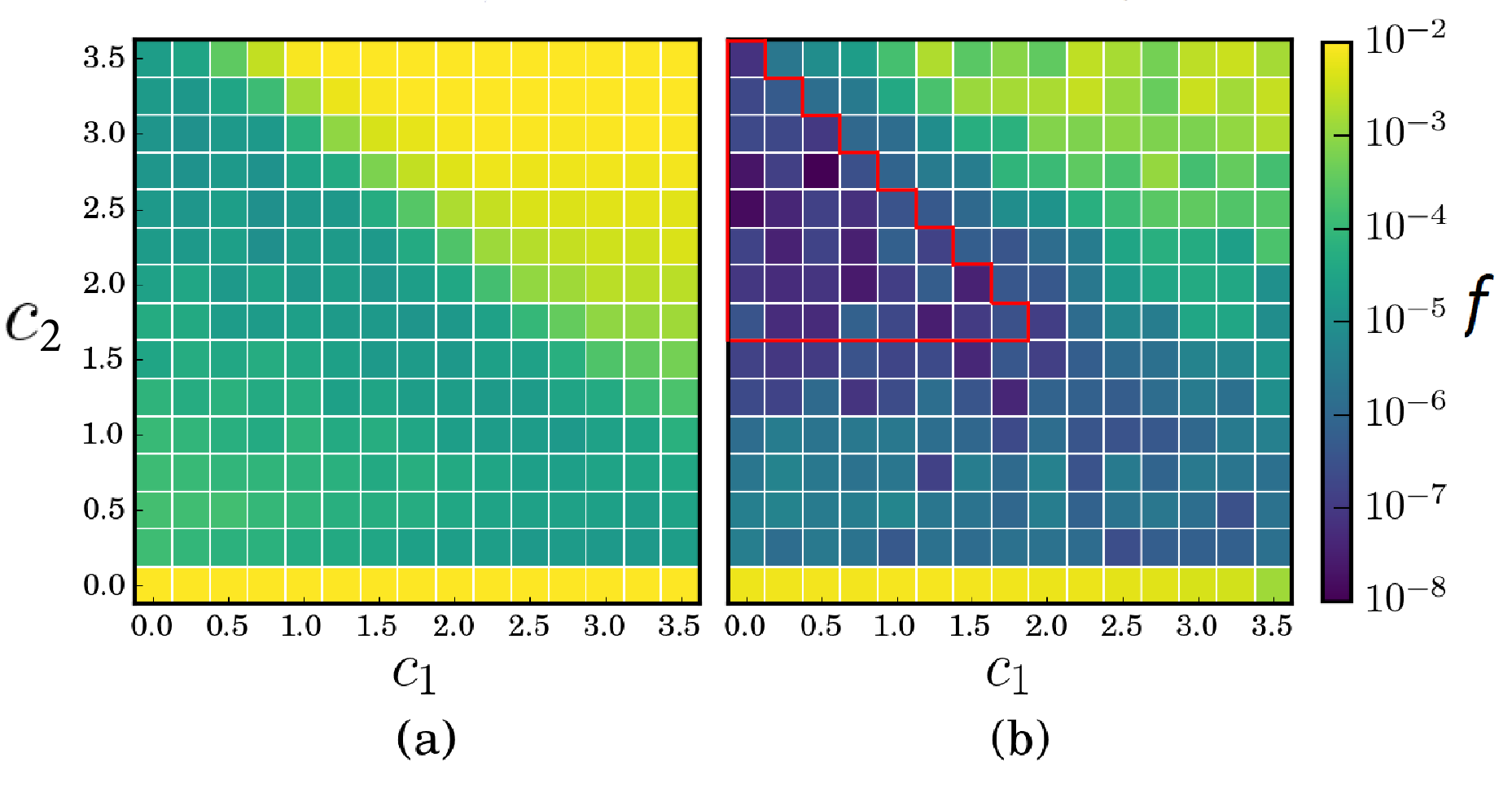}
	\caption{(color online)
		The mean objective function (a) and lowest objective function (b) for 100 realizations of a PSO search with $\omega$ = 0.25 as function of $c_1$ and $c_2$ indicate an optimal area in hyper-parameter space as indicated.}\label{hyperparam_opt}
\end{figure}

In PSO, the inertia $(w)$, cognition $(c_1)$ and social $(c_2)$ hyper-parameters need to be chosen
according to the underlying optimization problem, and we have performed a grid search method
to gain insight into their role. For each value of these, we run 100 instances of the PSO method and keep track of both the mean and lowest value of the objective function. In Fig.~\ref{hyperparam_opt} we show the results for $\omega=0.25$ and a range of $c_1$ and $c_2$ values. From this data we deduce the optimum hyperparameter subspace for $w=0.25$ as a triangular area in parameter space where $c_1 \geq 0.0$, $c_2 \geq 1.75$ and $c_1+c_2 \leq 3.50$. Similar studies for larger $\omega$ yield slightly smaller optimal subspaces, while significant lowering of $\omega$ also does not increase performance [49,50] --- we thus fix $w=0.25$ and keep $c_1 \geq 0.0$, $c_2 \geq 1.75$ and $c_1+c_2 \leq 3.50$.

\begin{figure}[t!]
	\includegraphics[width=0.90\linewidth]{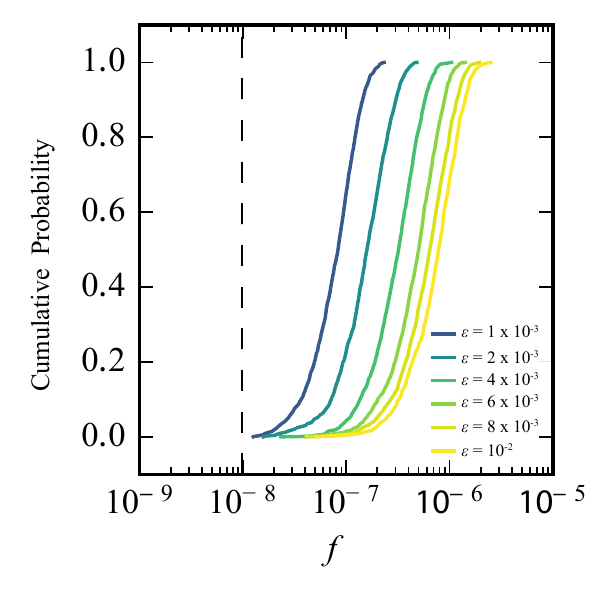}
	\caption{(color online)
		CDF's of $f(x_0 + \varepsilon dx_i)$ for $i$ ranging from 0 to 999,
		for values of $\varepsilon$ as indicated in the legend, and $x_0$ the solution shown in
		Fig.~2b.
		The objective function value $f(x_0)$ of the unperturbed solution is marked by the dashed black line.}\label{cumulative_good}
\end{figure}

\subsection{Local Minima Check}

\begin{figure}[h]
	\includegraphics[width=0.90\linewidth]{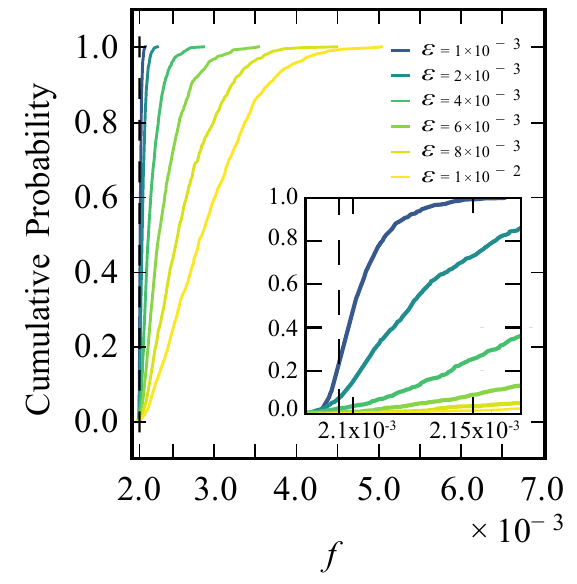}
	\caption{(color online)
		CDF's of $f(x_0 + \varepsilon dx_i)$ for $i$ ranging from 0 to 999,
		for values of $\varepsilon$ as indicated in the legend, and $x_0$ the solution with a much higher value of $f$ than in Fig.~2.
		The objective function value $f(x_0)$ of the unperturbed solution is marked by the dashed black line (see inset); a small fraction of perturbed solutions are seen to lower $f$, showing that $x_0$ is strict minimum.}\label{cumulative_bad}
\end{figure}

PSO discovers many realizations with very low objective function values.
To explore the objective function landscape, we sample the variation of the objective value in the vicinity of such solutions.
Specifically, we start from the final solution $x_0$
shown in Fig.~2b, generate 1000 random 24-dimensional vectors $dx_i$ with each entry uniformly distributed between -1 and 1, and
then calculate the cumulative distribution functions (CDFs) of $f(x_0 + \varepsilon dx_i)$ for $\varepsilon$ ranging from $10^{-3}$ to $10^{-2}$.
In all cases we find that $f(x_0 + \varepsilon dx_i) > f(x_0)$ (Fig.~\ref{cumulative_good})
While this is no proof that $x_0$ corresponds to a true local minimum, it strongly indicates that $x_0$ --- within an accuracy of $10^{-3}$ --- is a good solution of the PSO algorithm.
We note that a similar analysis for a solution with a much higher value of $f$
does not strictly satisfy $f(x_0 + \varepsilon dx_i) > f(x_0)$ (Fig.~\ref{cumulative_bad}).

\end{document}